\documentstyle[twocolumn,aps,prl,epsfig]{revtex}
\DeclareFontShape{OML}{cmm}{m}{b}{%
   <-> cmmib10}{}
\DeclareMathAlphabet{\mathbf}{OML}{cmm}{m}{b}
\DeclareSymbolFont{boldletters}{OML}{cmm}{m}{b}
\DeclareMathSymbol{\bsigma}{\mathord}{boldletters}{27} 

\def\case#1#2{\textstyle{#1\over#2}\displaystyle}
\def\i{{\rm i}}

\def\diag{{\rm diag}}

\begin{document}
\wideabs{\title{Magnetization Plateaus in a Solvable 3-Leg
Spin Ladder} 
\author{J. de Gier and  M.~T. Batchelor}
\address{Department of Mathematics, School of Mathematical Sciences,\\
Australian National University, Canberra ACT 0200, Australia}
\date{\today}
\draft
\maketitle
\begin{abstract}
We present a solvable ladder model which displays magnetization plateaus at
fractional values of the total magnetization. Plateau signatures are
also shown to exist along special lines. The model has isotropic 
Heisenberg interactions with additional many-body terms. The phase diagram 
can be calculated exactly for all values of the rung coupling and the 
magnetic field. We also derive the anomalous behaviour of the
susceptibility near the plateau boundaries. There is good agreement
with the phase diagram obtained recently for the pure Heisenberg
ladders by numerical and perturbative techniques.
\end{abstract}
}
\narrowtext

One of the surprising aspects of low-dimensional quantum systems with
long range finite interactions is the occurence of fractional magnetization
plateaus. Theoretical \cite{OYA,T,CHP,H,O} considerations have
revealed that such plateaus originate from interactions 
beyond those of nearest neighbours. 
Spin ladders provide natural examples of systems with non trivial
magnetization plateaus since they can be reformulated as spin 
chains with longer range interactions. They have the advantage
that they are accessible experimentally and spin ladders in a field
have attracted quite some attention over the past few years
\cite{Chab}. Fractional values of the total magnetization were
measured only very recently in 1D compounds \cite{Nar,Shir} and  in a
2D system \cite{Kag}. 

Groundstate phase diagrams and magnetization plat\-eaus have been
calculated for a special family of ladder models for which partially
exact results could be obtained \cite{HMT}. We consider an integrable
3-leg spin ladder, or spin tube for periodic boundary conditions
\cite{BM,dGBM}. Its phase diagram can be calculated for all
values of the rung coupling and the magnetic field from the Bethe
Ansatz solution. 
The spins along each leg and 
each rung have an isotropic Heisenberg interaction, with the introduction
of many-body terms to retain integrability.
The model is a generalisation of Wang's 2-leg ladder model \cite{W}.
The overall phase diagram of the 2- and 3-leg ladders compare well with 
those obtained recently for the pure Heisenberg ladders, by DMRG,
bosonization, series expansions and mappings to effective 
Hamiltonians (see, e.g., Refs.~\cite{CHP,M,TLPRS,COAIQa}).  

The Hamiltonian of our model is
\begin{equation}
H = \sum_{i=1}^L H^{\rm leg}_{i,i+1} + \sum_{i=1}^L H^{\rm
rung}_i + \sum_{i=1}^L H^{\rm field}_i ,
\label{eq:Ham}
\end{equation}
where, 
\begin{eqnarray}
H^{\rm leg}_{i,j} &=& \case{1}{8} \prod_{l=1}^3 \left( 1+
\bsigma_i^{(l)} \cdot \bsigma_{j}^{(l)}\right), \label{eq:hleg}\\
H^{\rm rung}_i &=& \sum_{l=1}^3 \case{1}{2} J_l \left(
\bsigma_i^{(l)} \cdot \bsigma_i^{(l+1)} -1 \right),\label{eq:hrung}\\
H^{\rm field}_i &=& - h \sum_{l=1}^3 
(\sigma^z)_i^{(l)}. \label{eq:hfield}
\end{eqnarray}
The operators $(\sigma^x)_i^{(l)}$, $(\sigma^y)_i^{(l)}$ and
$(\sigma^z)_i^{(l)}$ act as the corresponding Pauli matrices on the
$(i,l)$th factor in the Hilbert space.

As shown in \cite{dGBM}, the Hamiltonian (\ref{eq:Ham}) is integrable
for $h=0$. The addition of the magnetic field term however, does not
destroy integrability since $[H^{\rm leg}_{i,j}, H^{\rm field}_i +
H^{\rm field}_{j}]= 0$. In the following we set $J_1=J_2=J$ and
$J_3=J'$, so that we can go from the isotropic tube, $J'=J$, to the
ladder, $J'=0$.  

{}For $h=0$ it is convenient to change to the basis where the square and the
$z$-component of the total spin of a given triangle, ${\mathbf S}=
\bsigma^{(1)} + \bsigma^{(2)} + \bsigma^{(3)}$ are diagonal \cite{dGBM}. It
follows that the eight states on a given triangle fall into a
spin-$\frac{3}{2}$ quadruplet and two spin-$\frac{1}{2}$
doublets. We will denote these states by $|2s_z \rangle_q$ for the
quadruplet and $|2s_z \rangle_{d_i},\;(i=1,2)$ for each of the doublets.
Switching on the magnetic field breaks this symmetry further
due to the Zeeman splitting. The energies of the rung and field
Hamiltonians are given by
\begin{eqnarray}
&&H^{\rm rung} + H^{\rm field} = \diag \{ -3J - h,-3J + h,\nonumber\\
&&-2J'-J - h,-2J'-J + h, - 3h,- h, h, 3h\}, 
\end{eqnarray}
on the states 
\[\{|+\rangle_{d_1}, |-\rangle_{d_1},
|+\rangle_{d_2}, |-\rangle_{d_2}, |+3\rangle_q, |+\rangle_q,
|-\rangle_q, |-3\rangle_q\}.\]
Since ${\bf S}^2$ and $S^z$ commute with $H$, the total spin and its
$z$-component are good quantum numbers, as in Ref. \cite{HMT}.

$H$ can be diagonalized using the Bethe Ansatz. It is important to
note that the Hamiltonian (\ref{eq:hleg}) does not change under the
change of basis given above. Furthermore, (\ref{eq:hleg}) is invariant
under any choice of reference state (or pseudo-vacuum)
$|\Omega\rangle$ and any assignment of Bethe Ansatz pseudo
particles. For each choice however, one has to re-interpret this
assignment. The rung and field Hamiltonians do alter with the choice
of ordering, but the change is just a rearrangement of their
eigenvalues along the diagonal. We use this property to our advantage
by doing calculations with that choice of ordering for which the Bethe
Ansatz reference state is closest to the true groundstate of the
system. The eigenenergies of $\sum_{i=1}^L H^{\rm leg}_{i,i+1}$ are
given by  
\begin{equation}
E^{\rm leg} = -\sum_{j=1}^{M_1} \frac{1}{(\lambda_j^{(1)})^2 +
\frac{1}{4}},
\end{equation}
where the number $\lambda_j^{(1)}$ satisfy the well known
Bethe Ansatz equations \cite{S}, 
\begin{equation}
\left( \frac{\lambda_j^{(1)} - \frac{\i}{2}}{\lambda_j^{(1)} +
\frac{\i}{2}} \right)^L = \prod_{k \neq j}^{M_1}
\frac{\lambda_j^{(1)} - \lambda_k^{(1)} - \i}{\lambda_j^{(1)} -
\lambda_k^{(1)} + \i} \prod_{k=1}^{M_2}
\frac{\lambda_j^{(1)} - \lambda_k^{(2)} + \frac{\i}{2}}{\lambda_j^{(1)} -
\lambda_k^{(2)} - \frac{\i}{2}}, \label{eq:BAE}
\end{equation}
and for $r=2,\ldots,7$ with $M_8=0$,
\begin{eqnarray}
&&\prod_{k \neq j}^{M_r}
\frac{\lambda_j^{(r)} - \lambda_k^{(r)} - \i}{\lambda_j^{(r)} -
\lambda_k^{(r)} + \i} = \nonumber\\
&&\prod_{k=1}^{M_{r-1}}
\frac{\lambda_j^{(r)} - \lambda_k^{(r-1)} - \frac{\i}{2}}{\lambda_j^{(r)} -
\lambda_k^{(r-1)} + \frac{\i}{2}} \prod_{k=1}^{M_{r+1}}
\frac{\lambda_j^{(r)} - \lambda_k^{(r+1)} - \frac{\i}{2}}{\lambda_j^{(r)} -
\lambda_k^{(r+1)} + \frac{\i}{2}},
\end{eqnarray}
where $j=1,\ldots,M_r$. 

In the following we will restrict ourselves to the quadrant $J,h \geq
0$ and consider the magnetization, which is defined by
\begin{equation}
M = \frac{1}{n L} \sum_{i=1}^L \sum_{l=1}^n \left( \sigma^z)_i^{(l)}
\right),
\end{equation}
where $n$ is the number of legs.

\section*{The 2-leg ladder}

In this section we briefly review and expand the results of Wang \cite{W}
before treating the 3-leg case. For the 2-leg case the rung states fall
into a singlet and a triplet. The phase diagram in the $J,h>0$
quadrant is determined by the competition between the singlet state
and the spin up state of the triplet. The difference between their
respective energies changes sign at $h=J$. This line therefore divides
phase space into two regions. In each of these regions a convenient
choice of Bethe Ansatz reference state and pseudo particles may be
made. The value of the gaps can then easily be calculated and it
follows that there is a massive phase for $h-J > 2$ where the
groundstate is the simple product of the spin up triplet state. In
this phase $\langle M\rangle =1$. For $J-h > 2$ there is another massive
phase where the groundstate consists of singlets on each rung. Here
evidently $\langle M\rangle =0$. In between these two phases lies a
massless phase where the magnetization varies continuously. On
approaching the lines $h=J \pm 2$ from within the massless phase, the
susceptibility shows the familiar square root singularity \cite{KDNPT}.

On the line $h=J$ the spin up triplet and the singlet state are
degenerate. Therefore, on this line the magnetization $\langle M\rangle
=\case{1}{2}$. At the point $J=h=\log 2$ the other excitations become
massless and a completely massless phase is entered. This phase
actually covers a finite region around the origin which seems to be
common to this type of solvable ladder model. The other point that can
be calculated exactly marking its phase boundary is $J=2,h=0$. The
line $h=J$ can be regarded as the onset of a plateau
boundary. Although there is no singularity in the magnetic
susceptibility in the present case, it will appear as soon as the
plateau opens. It is expected that the opening of this plateau is
governed by anisotropy \cite{H,O}. 

\begin{figure}[htb]
\setlength{\unitlength}{1mm}
\begin{picture}(70,60)
\put(0,0){\epsfig{height=6cm,width=7cm,file=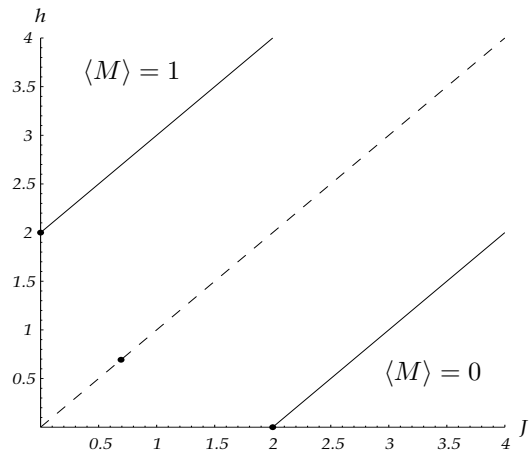}}
\put(10,50){$\langle M\rangle =1$}
\put(50,10){$\langle M\rangle =0$}
\end{picture}
\vspace{4mm}
\caption{The solvable 2-leg ladder phase diagram. The dashed line 
$h=J$ divides phase space into two regions. The bold lines
given by $h=J\pm2$ are phase boundaries.}  
\label{fig:2-legladder}
\end{figure}

\section*{The tube: $J'=J$}

Due to the extra symmetry in this case, the two doublets are
degenerate for the rung Hamiltonian and only three eigenvalues of the
Hamiltonian $H^{\rm rung}_i + H^{\rm field}_i$ are competing,
$-(3J+h)$, $-(3J-h)$ and $-3h$. Their differences change sign at the
lines,  $2h=3J$ and $4h=3J$, dividing the phase space in three
regions. In each of these regions we can make a convenient choice of
ordering the states to facilitate our calculations. 

{\em i)} $h \geq \case{3}{2} J$.
\label{ss:tube1}
We choose our states to be ordered in increasing energy with respect
to $h^{\rm rung}+h^{\rm field}$, i.e. $\{|+3\rangle_q, |+\rangle_{d_1},
|+\rangle_{d_2},\ldots\}$. The energy, up to an irrelevant constant,
is given by 
\begin{eqnarray}
E&=& 
- \sum_{j=1}^{M_1} \left( \frac{1}{(\lambda^{(1)}_j)^2 +
\case{1}{4}} - (2h-3J) \right) \nonumber\\
&&{}+ 3JM_3 + (2h-3J)M_4 +3JM_6 +2h M_7.
\end{eqnarray}
{}From this expression it follows straightforwardly that for $2h-3J > 4$
we have $M_1=0 $ for the groundstate energy. In this phase, the pseudo
vacuum is the true groundstate and the total magnetization $\langle
M\rangle =1$ and all excitations are gapped. Below the line
$h=\case{3}{2} J+2$, a finite density of $|+\rangle_{d_1}$ and
$|+\rangle_{d_2}$ states appear in the groundstate. This phase is
massless and the mean magnetization varies continuously. As the
magnetization reaches its maximum value, the susceptibility diverges
as,
\begin{equation}
\chi \sim (\case{3}{2}J+2-h)^{-\frac{1}{2}}.
\label{eq:chi}
\end{equation}
Upon reaching the line $h=\case{3}{2}J$, the three states have become
degenerate in energy while all other excitations remain massive for
large enough $J$. On this line the magnetization is $\langle M \rangle
=\case{5}{9}$ \cite{COAIQb}. This line can be regarded as a plateau
boundary in a similar way as the line $h=J$ for the 2-leg ladder. The
plateau will open upon introduction of anisotropy.

The first excitations to become massless on this line are $|+\rangle_q$,
$|-\rangle_{d_1}$ and $|-\rangle_{d_2}$. This happens at the point $3J
= \pi/\sqrt{3} - \log 3$. 

{\em ii)} $\case{3}{4}J \leq h \leq \case{3}{2}J$.
\label{ss:tube2}
Here we choose our state to be ordered as $\{|+\rangle_{d_1},
|+\rangle_{d_2}, |+3\rangle_q, \ldots\}$. The energy in this case
takes the form, 
\begin{eqnarray}
E &=& 
- \sum_{j=1}^{M_1}
\frac{1}{(\lambda^{(1)}_j)^2 + \case{1}{4}} + (3J-2h)M_2 + (4h-3J)M_3
\nonumber\\ 
&&{} + (3J-2h)M_5 + 2h (M_6 + M_7). 
\end{eqnarray}
Following the same reasoning as above, we find that for $3J-2h$ large
enough $M_2=0$ for the groundstate. In this region the magnetization
is $\langle M\rangle =\case{1}{3}$ and every {\em magnetic} excitation is
massive. Because of the degeneracy of the spin up states of both
doublets, there are massless excitations because the groundstate is
essentially that of a spin-$\case{1}{2}$ chain. The $|+3\rangle_q$
excitation becomes massless at $3J-2h = 2\log 2$ where the massless
phase is entered. Here $\langle M\rangle$ may vary continuously. The
susceptibility shows the square root singularity upon
approaching the plateau from within this phase. 

{\em iii)} $h \leq \case{3}{4} J$.
Here, the ordering for the lowest energy rung states is given by
$\{|+\rangle_{d_1}, |+\rangle_{d_2}, |-\rangle_{d_1},
|-\rangle_{d_2},\ldots \}$. The energy in this case takes the
form
\begin{eqnarray}
E &=& 
- \sum_{j=1}^{M_1}
\frac{1}{(\lambda^{(1)}_j)^2 + \case{1}{4}} + 2h M_2 \nonumber\\
&&{}+ (3J-4h)M_4 + 2h (M_5 + M_6 + M_7). 
\end{eqnarray}
Above the line $h=\log 2$ there is the plateau phase with $\langle
M\rangle =\case{1}{3}$. Below this line, the excitations
$|-\rangle_{d_1}$ and $|-\rangle_{d_2}$ become massless and the
magnetization is allowed to vary continuously. On the line $h=0$ the
two doublets are degenerate and the magnetization $\langle M\rangle
=0$. As before, we may regard the line $h=0$ as a plateau boundary. 

\begin{figure}[htb]
\setlength{\unitlength}{1mm}
\begin{picture}(70,60)
\put(0,0){\epsfig{height=6cm,width=7cm,file=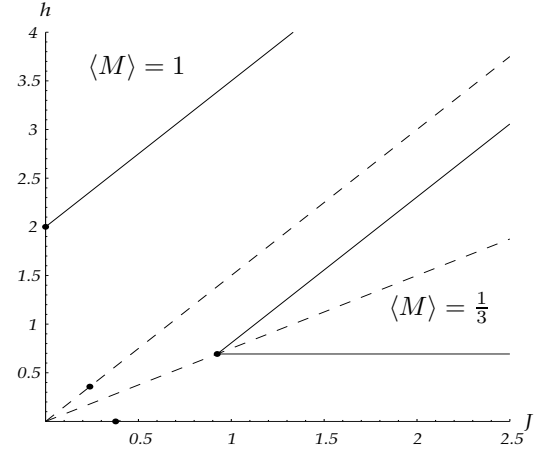}}
\put(10,50){$\langle M\rangle =1$}
\put(50,18){$\langle M\rangle =\case{1}{3}$}
\end{picture}
\vspace{4mm}
\caption{The solvable 3-leg tube phase diagram. The dashed lines 
$2h=3J$ and $4h=3J$ divide phase space into three regions. 
The bold lines $h=\case{3}{2}J+2$, $h=\case{3}{2}J-\log 2$
and $h=\log 2$ are phase boundaries.} 
\label{fig:phasetube}
\end{figure}

Our results for the tube are summarised in Fig.~\ref{fig:phasetube}.

\section*{The ladder: $J'=0$}

{}For the ladder the degeneracies are lifted and the massless regions
display more structure. Phase space is now divided into six regions
due to level crossings, of which the boundaries are given by
$h=\case{k}{4}J$ for $k=1,2,3,4,6$. The only boundaries where
level crossings concerning the groundstate occur are those with $k=3$
and $k=6$, i.e. the same as those for the tube. A similar analysis as
for the tube gives the following results.

{\em i)} $h \geq \case{3}{2} J$.
\label{ss:ladder1}
As in the case for the tube, for strong magnetic fields the ladder is
completely magnetized. Below the line $h=\case{3}{2}J + 2$, some of
the $|+\rangle_{d_1}$ states appear and the magnetization may
vary continuously. Again, the susceptibility diverges with a square
root singularity as this plateau is approached. On the line $2h=3J$,
the states $|+3\rangle_q$ and $|+\rangle_{d_1}$ are degenerate and the
magnetization thus is given by $\langle M\rangle =\case{2}{3}$. As
before this line is interpreted as a plateau boundary. The line does
not extend to the origin, but at some finite value of $J=\log 2$ other
magnetic excitations become massless. 

{\em ii)} $\case{3}{4}J \leq h \leq \case{3}{2}J$.
{}For $3J-2h$ large enough only the state $|+\rangle_{d_1}$ appears in
the groundstate. In this region the magnetization is $\langle M\rangle
=\case{1}{3}$ and every excitation is massive. As opposed to the tube, the
two doublets are not degenerate and the groundstate is completely
polarized, as in Ref. \cite{HMT}. The $|+3\rangle_q$ excitation
becomes massless at $3J-2h = 2$ where the massless phase of the
previous paragraph is entered. Here $\langle M\rangle$ may very
continuously but the phase is still polarized. Also here the square
root singularity shows up when approaching this plateau. 

On the line $4h=3J$, the $\langle M\rangle=\case{1}{3}$ plateau extends into
the massless phase. On this line the states $|+3\rangle_q$ and
$|-\rangle_{d_1}$ are degenerate so that they may combine to a net
nonmagnetic excitation. It ends at some finite value of $J$ which
cannot be calculated analytically.  

{\em iii)} $h \leq \case{3}{4} J$.
Here, the ordering for the lowest energy rung states is given by
$\{|+\rangle_{d_1}, |-\rangle_{d_1}, |+\rangle_{d_2},
|-\rangle_{d_2},\ldots \}$. Above the line $h=2$ there is the plateau
phase with $\langle M\rangle =\case{1}{3}$. Below this line, the excitation
$|-\rangle_{d_1}$ becomes massless and the magnetization is allowed to
vary continuously. For $J$ not too small, the phase remains polarized.
The line $h=0$, where the magnetization $\langle M\rangle =0$,
may again be regarded as a plateau boundary. The second doublet enters
the groundstate at $J=\log 2$ \cite{dGBM} and the polarized excitations
become massless. As before, the susceptibility shows the square root
singularity upon approaching this line from above and we may regard
the line $h=0$ as a plateau boundary. 

\begin{figure}[htb]
\setlength{\unitlength}{1mm}
\begin{picture}(70,60)
\put(0,0){\epsfig{height=6cm,width=7cm,file=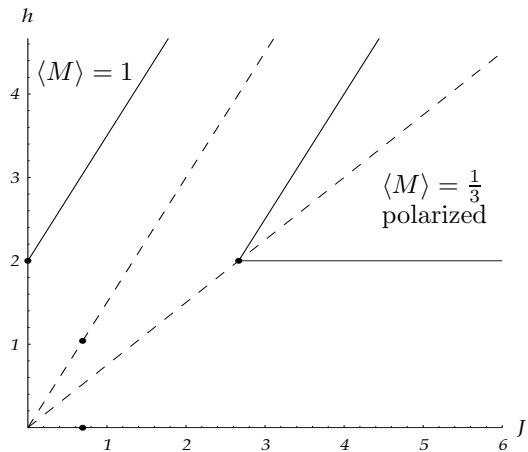}}
\put(4,50){$\langle M\rangle =1$}
\put(50,35){$\langle M\rangle =\case{1}{3}$}
\put(50,31){polarized}
\end{picture}
\vspace{4mm}
\caption{The solvable 3-leg ladder phase diagram. The dashed lines 
$2h=3J$ and $4h=3J$ divide phase space into three regions. The bold 
lines $h=\case{3}{2}J \pm 2$ and $h=2$  
are phase boundaries.}
\label{fig:phaseladder}
\end{figure}

Our results for the tube are summarised in Fig.~\ref{fig:phaseladder}.

\section*{Conclusion}

In summary we have considered a simple solvable model that displays
fractional magnetization plateaus. Such plateaus have been studied
theoretically only by perturbative or numerical techniques.  
The phase diagram and in particular the location of the plateaus and
the expected square root singularity have been calculated exactly.
The phase diagrams in
Figs.~\ref{fig:2-legladder}-\ref{fig:phaseladder} are to be
compared with those of the pure Heisenberg ladders given in
Ref.~\cite{CHP}. 
The overall agreement is seen to be excellent and our results should
provide a useful benchmark for further studies. 
Consideration of the complete eigenspectrum will allow the exact study
of thermodynamic effects. 
More intricate phase diagrams can be obtained similarly for solvable
$n$-leg ladder and tube models. 

In future work we hope to be able to include anisotropy in order to 
tune the width of the plateaus, see, e.g., the discussion
in Ref. \cite{CHP} with respect to the XXZ chain. 
For example, in this way the lines with magnetization $\langle M
\rangle = 0$, $\langle M \rangle = \case{2}{3}$ and $\langle M \rangle
= \case{5}{9}$ may disappear or may broaden to reveal a finite gap. The
fact that the models in this Letter map onto an {\em isotropic} low
energy effective Hamiltonian in a strong coupling approach
\cite{M} is consistent with this picture.  

This work has been supported by the Australian Research Council.


\begin{references}
\bibitem{COAIQb} A possible plateau at this value, i.e. $\langle
S^z\rangle = \case{5}{6}$, was discussed in Ref. \cite{COAIQa}.
\end{references}
\end{document}